\documentclass[a4paper,11pt]{iopart}
\usepackage[utf8]{inputenc}

\usepackage[british,english]{babel}
\usepackage[sorting=none,    
            backend=bibtexu, 
            style=numeric-comp,
            maxbibnames=99,
            firstinits=true,
            useprefix=true
                         ]{biblatex}%
\usepackage{csquotes}
\renewcommand{\revsdnamepunct}{\addspace}

\DeclareNameAlias{default}{last-first}

 \DeclareFieldFormat[article]{pages}{\mkfirstpage{#1}}

\DeclareBibliographyDriver{article}{%
	\usebibmacro{author},%
\addspace\thefield{year}\addspace\textsl{\thefield{journaltitle}}\addspace%
\iffieldundef{doi}{
	\iffieldundef{url}{
		\iffieldundef{volume}{\thefield{pages}}{{\bf\printfield{volume}}\addspace\thefield{pages}}%
	}{
		\href{\thefield{url}}{%
			\iffieldundef{volume}{\thefield{pages}}{{\bf\printfield{volume}}\addspace\thefield{pages}}%
		}%
	}%
}{
	\href{http://dx.doi.org/\thefield{doi}}{%
		\iffieldundef{volume}{\thefield{pages}}{{\bf\printfield{volume}}\addspace\printfield{pages}}%
	}%
}

}

\DeclareBibliographyDriver{book}{%
	\usebibmacro{author/editor+others/translator+others},%
\addspace\thefield{year}\addspace\textsl{\thefield{title}}%
\iffieldundef{edition}{}{%
	\addspace\printfield{edition},}
\addspace(\printlist{location}:\addspace\printlist{publisher})%
}

\usepackage{iopams}
\usepackage[colorlinks=true,linkcolor=blue,citecolor=blue,filecolor=cyan,urlcolor=blue,breaklinks=true]{hyperref}
\renewcommand*{\bibfont}{\footnotesize}
\usepackage{graphicx}

\begin{document}

\title[Stochastic thermodynamics of bipartite systems]{Stochastic thermodynamics of bipartite systems: transfer entropy inequalities and\\
 a Maxwell's demon interpretation}
\author{D Hartich, A C Barato, and U Seifert}
\address{II. Institut f\"ur Theoretische Physik, Universit\"at Stuttgart\\ 
Stuttgart 70550, Germany\\
}

\ead{hartich@theo2.physik.uni-stuttgart.de}

\def\ex#1{\langle #1 \rangle}

\begin{abstract}
We consider the stationary state of a Markov process on 
a bipartite system from the perspective of stochastic
thermodynamics. One subsystem is used to extract work 
from a heat bath while being affected by the
second subsystem. We show that the latter allows for a
transparent and thermodynamically consistent interpretation
of a Maxwell's demon. Moreover, we obtain an integral fluctuation 
theorem involving the transfer entropy from one subsystem 
to the other. 
Comparing three different inequalities, 
we show that the entropy decrease of the first subsystem 
provides a tighter bound on the rate of extracted work 
than both the rate of transfer entropy from this subsystem 
to the demon and the heat dissipated through the dynamics 
of the demon. The latter two rates cannot be ordered by an 
inequality as shown with the illustrative example of a 
four state system.

\end{abstract}

\section{Introduction}

Thermodynamics of information processing started a long time ago with a thought experiment about ``violations'' of the second law achieved by Maxwell's demon \cite{maxw01}. Among the seminal 
contributions to this field (see \cite{leff03} for a collection of papers) are Szilard's engine \cite{szil29}, Landauer's principle \cite{land61} and Bennett's work \cite{benn82}.

More broadly, access to small systems where fluctuations are not negligible is now possible and understanding the relation between thermodynamics and information
has become a problem of practical interest. For example, experimental  verifications of the conversion of 
information into work \cite{toya10a} and of Landauer's principle \cite{beru12} have been realized. 
Moreover, considerable theoretical progress has been made recently with the derivation of second law inequalities \cite{touc00,touc04,cao09,espo11} and  fluctuation relations 
\cite{saga10,ponm10,horo10,abre11a,kund12,saga12,saga12b} for feedback driven systems. The study of simple models has also played an important role \cite{cao04,horo11,horo11a,gran11,abre11,baue12,
kish12,espo12b,stra13,horo13,gran13,dian13a,espo13,andr08,andr13}. Particularly, 
Mandal and Jarzynski \cite{mand12} (see also \cite{mand13,bara13,bara14,deff13}) have introduced a model which clearly demonstrates 
 an idea expressed by Bennett \cite{benn82}: a tape can be used to do work as it randomizes itself.    

In related work \cite{bara13b,bara13a}, we have studied the relation between the rate of mutual information and the thermodynamic entropy production in bipartite systems. By bipartite systems
we mean a Markov process with states that are determined by two variables such that in a transition between states only one of the variables can change. We have obtained 
an analytical upper bound on the rate of mutual information and developed a numerical method to estimate Shannon entropy rates of continuous time series \cite{bara13a}. 

In the present paper we take the view that a bipartite system provides a simple and convenient description of a Maxwell's demon. More precisely, considering a
subsystem $y$ as a Maxwell's demon, we show that work extraction by the other subsystem $x$ leads to an entropy decrease of the external medium, with this entropy decrease being bounded by 
the entropy reduction of $x$ due to its coupling with $y$. As an advantage of this Maxwell's demon realization, the full thermodynamic cost is easily accessible,
being given by the standard entropy production of the bipartite system.   

Moreover, we also study transfer entropy within our setup. Transfer entropy is an informational theoretical measure of how the dynamics of a process depends on another process \cite{schr00}, being
an important concept in the analysis of time series \cite{hlav07}. Ito and Sagawa \cite{ito13} have recently obtained a fluctuation relation for very general dynamics
involving the entropy variation of the external medium due to a subsystem and transfer entropy. Here, we obtain a similar fluctuation relation for a bipartite system. This
fluctuation relation implies that the entropy decrease of the external medium due to subsystem $x$ is bounded by the transfer entropy from $x$ to $y$, where $y$
can be interpreted as a Maxwell's demon. 
  
Hence, we find three different bounds for the entropy decrease of the external medium due to $x$: the entropy reduction of $x$, the transfer entropy from $x$ to $y$ and the entropy increase
of the external medium due to $y$. We show that the entropy reduction of $x$ is always the best bound. Furthermore, studying a particular four state model we observe that the transfer entropy
can be larger than the entropy increase of the medium due to $y$.

The paper is organized as follows. In the next section, we define bipartite systems and the thermodynamic entropy production. Moreover, we explain in which sense a subsystem can be interpreted as a Maxwell's demon. 
We define the transfer entropy and obtain an analytical upper bound for it in Sec. \ref{sec3}. In Sec. \ref{sec4} we prove an integral fluctuation relation involving the transfer entropy and  show that the transfer entropy from $x$ to $y$ is larger than
the entropy reduction of $x$ due to $y$. Our results are illustrated with a simple four state system in Sec. \ref{sec5}, where we summarize and compare the different inequalities obtained in this paper.
We conclude in Sec. \ref{sec6}.


\section{Bipartite systems and thermodynamic entropy production}
\label{sec2}

\subsection{Basic definitions and inequalities}
We restrict to a class of Markov processes which we call bipartite \cite{bara13b,bara13a}. The states are labeled by the pair of variables $(\alpha,i)$, where $\alpha\in\{1,\ldots,\Omega_x\}$ and $i\in\{1,\ldots,\Omega_y\}$.
The transition rates from $(\alpha,i)$ to $(\beta,j)$,  are given by
\begin{equation}	
w_{ij}^{\alpha\beta}\equiv\left\{
\begin{array}{ll} 
 w^{\alpha\beta}_i & \quad \textrm{if $i=j$ and $\alpha\neq\beta$}, \\
 w^{\alpha}_{ij} & \quad  \textrm{if $i\neq j$ and $\alpha=\beta$},\\
 0 & \quad \textrm{if $i\neq j$ and $\alpha\neq\beta$}. 
\end{array}\right.\,
\label{defrates2}
\end{equation}
The central feature of the network of states is that when a jump occurs only one of the variables changes. Examples of a bipartite systems, {\sl inter alia}, are stochastic models for cellular sensing \cite{bara13b,lan12,meht12}, where one 
variable could represent the activity of a receptor and the other the concentration of some phosphorylated internal protein. We also denote a state of the system at time $t$ 
by $z(t)=(x(t),y(t))$, where $x(t)\in\{1,\ldots,\Omega_x\}$ and $y(t)\in\{1,\ldots,\Omega_y\}$. Hence, the subsystem $x$ is related to the variable denoted by Greek letters and the subsystem $y$  is
related to the Roman letters.

The rate of entropy increase of the external medium \cite{seif12} 
is then divided into two parts, one caused by jumps in the $x$ variable and the other by jumps in the $y$ variable. More precisely, we define
\begin{equation}
\sigma_x\equiv \sum_{i,\alpha}P_i^\alpha \sum_{\beta\neq\alpha} w^{\alpha\beta}_i\ln \frac{w^{\alpha\beta}_i}{w^{\beta\alpha}_i}
\label{sdef2}
\end{equation} 
and
\begin{equation}
\sigma_y\equiv \sum_{i,\alpha}P_i^\alpha\sum_{j\neq i} w^\alpha_{ij}\ln\frac{w^\alpha_{ij}}{w^\alpha_{ji}},
\label{sdef3}
\end{equation} 
where $P_i^{\alpha}$ is the stationary probability distribution. The total entropy production, which fulfills the second law of thermodynamics, is then given by \cite{seif12}
\begin{equation}
\sigma\equiv \sigma_x+\sigma_y\ge 0,
\label{sdef1}
\end{equation} 
The rate $\sigma_x$ ($\sigma_y$) can be interpreted as the rate of increase of the entropy of the external medium due to the dynamics of the $x$ ($y$) subsystem. It is important 
to notice that $\sigma_x$ is not a coarse grained entropy rate \cite{espo12,mehl12,rold10}: knowing only the 
$x$ time series is not sufficient to calculate $\sigma_x$.

The rate of change of the Shannon entropy of the system is known to be zero in the stationary state, this can be written as \cite{seif12}
\begin{equation}
\sum_{i,\alpha}P_i^\alpha \sum_{\beta\neq\alpha} w^{\alpha\beta}_i\ln \frac{P^{\alpha}_i}{P^{\beta}_i}+\sum_{i,\alpha}P_i^\alpha \sum_{j\neq i} w^{\alpha}_{ij}\ln \frac{P^{\alpha}_i}{P^{\alpha}_j}=0.
\label{sys01}
\end{equation}
Considering the term originating due to the $x$ jumps we define 
\begin{equation}
h_x\equiv \sum_{i,\alpha}P_i^\alpha \sum_{\beta\neq\alpha} w^{\alpha\beta}_i\ln \frac{P^{\alpha}_i}{P^{\beta}_i}. 
\label{eq:hx_def}
\end{equation}
We interpret this quantity as the rate at which the entropy of the subsystem $x$ is reduced due to its coupling with $y$. This can be understood in the following way. If the state of the
subsystem $y$ is $i$, then the stationary probability of state $\alpha$ given $i$ is $P(\alpha|i)= P_i^\alpha/\sum_\beta P_i^\beta$. Therefore, the rate of change of the Shannon entropy
of the subsystem $x$ for $y=i$ is just $\sum_{\alpha}P_i^\alpha \sum_{\beta\neq\alpha} w^{\alpha\beta}_i\ln \frac{P(\alpha|i)}{P(\beta|i)}$. Summing over all possible $y$ we obtain (\ref{eq:hx_def}).
In this view it is as if the subsystem $x$'s transition rates $w_i^{\alpha\beta}$ and probabilities $P(\alpha|i)$ depend on time due to the $y$ jumps. In \ref{a1} we consider a functional of the stochastic trajectory which when averaged
gives $h_x$. With this functional, this interpretation of $h_x$ becomes even more clear. Similarly, for the subsystem $y$ we have
\begin{equation}
h_y\equiv \sum_{i,\alpha}P_i^\alpha \sum_{j\neq i} w^{\alpha}_{ij}\ln \frac{P^{\alpha}_i}{P^{\alpha}_j}. 
\label{eq:hy_def}
\end{equation}  
From (\ref{sys01}) it follows 
\begin{equation}
h_x=-h_y,
\label{sys0}
\end{equation}
i.e., the entropy reduction of the subsystem $x$ equals the entropy increase of subsystem $y$.

Besides the second law inequality for the full system (\ref{sdef1}), we also have for the total entropy production caused by transitions in the subsystem $x$   
\begin{equation}
\sigma_x+h_x= \sum_{i,\alpha}P_i^\alpha \sum_{\beta\neq\alpha} w^{\alpha\beta}_i\ln \frac{w^{\alpha\beta}_iP_i^\alpha}{w^{\beta\alpha}_iP_i^\beta}\ge 0.
\label{totx}
\end{equation}  
This inequality has been considered explicitly in \cite{dian13} and is a direct consequence of the log sum inequality. In \ref{a1} we prove a more general integral fluctuation relation
which implies (\ref{totx}). This  fluctuation relation  is similar to the fluctuation relation for the house-keeping entropy obtained by considering the dual dynamics \cite{seif12,spec05a}.
The same inequality is valid for the subsystem $y$,
\begin{equation}
\sigma_y+h_y=  \sum_{i,\alpha}P_i^\alpha\sum_{j\neq i} w^\alpha_{ij}\ln\frac{w^\alpha_{ij}P_i^\alpha}{w^\alpha_{ji}P_j^\alpha}\ge 0.
\label{toty}
\end{equation}  

\subsection{Subsystem $y$ as a Maxwell's demon}

Let us now consider a case where $\sigma_x$ is negative, i.e., the entropy of the external medium decreases at rate $-\sigma_x$ due to the subsystem $x$ dynamics. 
From relations (\ref{sys0}), (\ref{totx}), and (\ref{toty}) we obtain the following inequalities 
\begin{equation}
\sigma_y\ge h_x \ge -\sigma_x.
\label{eq:sx_hy_sy}
\end{equation}    
The second inequality can be interpreted in the following way.  If we consider the subsystem $y$ as a Maxwell's demon, then the rate of entropy reduction of the external medium $-\sigma_x$ is bounded by 
the rate $h_x$ at which the entropy of the subsystem $x$ is reduced due to its coupling to the Maxwell's demon. Furthermore, the first inequality contains an integrated description with the rate at which entropy
increases in the Maxwell's demon $-h_y=h_x$ being bounded by the rate of entropy increase in the external medium due to the $y$ dynamics $\sigma_y$.

For a more specific interpretation involving the first law we assume that the transition rates take the following local detailed balance form
\begin{equation}
\ln \frac{w_i^{\alpha\beta}}{w_i^{\beta\alpha}}= (E_i^\alpha-E_i^\beta)-\omega_i^{\alpha\beta},
\end{equation} 
where $E_i^\alpha$ is the internal energy of the state $(\alpha,i)$ and $\omega_i^{\alpha\beta}$ is the work extracted from the system in the jump $(\alpha,i)\to (\beta,j)$. We set $k_B T=1$ throughout the paper.  
In the stationary state the rate of internal energy change  due to $x$ jumps is given by  
\begin{equation}
\epsilon_x= \sum_{i,\alpha,\beta}P_i^{\alpha}w_i^{\alpha\beta}(E_i^\beta-E_i^\alpha).
\end{equation}
The rate of extracted work is
\begin{equation}
\omega^{{\rm out}}= \sum_{i,\alpha,\beta}P_i^{\alpha}w_i^{\alpha\beta}\omega_i^{\alpha\beta}.
\end{equation}
From the relation $\sigma_x= -\epsilon_x-\omega^{{\rm out}}$ and the first law we identify $\sigma_x$ as the dissipated heat due to the $x$ jumps. Likewise $\sigma_y$ is the dissipated heat due to $y$ jumps. 
For the special case $\epsilon_x=0$, we have $-\sigma_x=\omega^{{\rm out}}$: the second inequality in (\ref{eq:sx_hy_sy}) implies that the rate of extracted work is bounded by $h_x$. 
The first inequality in (\ref{eq:sx_hy_sy}) means that
the rate of entropy decrease $h_x$ is bounded by the heat $\sigma_y$ that is dissipated by the demon.
 
Let us make a comparison with the model introduced by Mandal and Jarzynski \cite{mand12}, where a tape composed of bits interacts with a system connected to a heat bath. By
increasing the Shannon entropy of the tape the system can deliver work to a work reservoir. In the above interpretation, this delivered work corresponds to $-\sigma_x$ and the tape
is analogous to the subsystem $y$: the subsystem $x$ can deliver work by increasing the entropy of the subsystem $y$. In this sense we can see the subsystem $y$ as an information or
entropy reservoir (see \cite{bara14,deff13} for definitions of an information reservoir). 

 
Summarizing the above discussion, bipartite systems provide a particularly transparent description of Maxwell's demon, with
the full thermodynamic cost being easily accessible through the standard second law inequality (\ref{sdef1}). We proceed by defining transfer entropy, which, as we will show in Sec. \ref{sec4},
also provides a bound for $-\sigma_x$.

\section{Shannon entropy rate and transfer entropy}
\label{sec3}

We first consider a discrete time Markov chain with time spacing $\tau$ and transition probabilities corresponding to the transition rates (\ref{defrates2}), i.e.,  
\begin{equation}
W_{ij}^{\alpha\beta}\equiv
\cases{
 w^{\alpha\beta}_i\tau   &\textrm{if $i=j$ and $\alpha\neq\beta$}, \\
 w^{\alpha}_{ij}\tau   &\textrm{if $i\neq j$ and $\alpha=\beta$},\\
 0   &\textrm{if $i\neq j$ and $\alpha\neq\beta$},\\
 1-\sum_{k\neq i} w^{\alpha}_{ik}\tau-\sum_{\gamma\neq\alpha}w^{\alpha\gamma}_{i}\tau   &\textrm{if $i=j$ and $\alpha=\beta$}. 
}
\label{defrates}
\end{equation}
We denote the full state of the system at time $n\tau$ by $z_n= (x_n,y_n)$, where $x_n\in\{1,\ldots,\Omega_x\}$ and $y_n\in\{1,\ldots,\Omega_y\}$. For the case where $x_{n-1}=\alpha$, $x_{n}=\beta$, $y_{n-1}=i$, and $y_{n}=j$,
we represent the transition probability $W_{ij}^{\alpha\beta}$ by $W(x_n,y_n|x_{n-1},y_{n-1})$. Furthermore, the stochastic trajectory of the full process is written as
$z_0^n= (z_0,z_1,\ldots,z_n)$. Whereas $z_0^n$ is Markovian, the stochastic trajectories of the two coarse grained processes $x_0^n$ and $y_0^n$ are in general non-Markovian.  

The Shannon entropy rate is a measure of how much the Shannon entropy of a stochastic trajectory increases as we increase the length of the trajectory $n$. For a generic process $a_0^n$ (where $a=x,y,z$) it is defined as
\begin{equation}
H_{a}\equiv -\lim_{n\to \infty}\frac{1}{n\tau}\sum_{a_0^n}P[a_0^n]\ln P[a_0^n],
\label{entrate}
\end{equation}
where $P[a_0^n]$ is the probability of the trajectory $a_0^n$. Particularly, since the full process is Markovian, its Shannon entropy rate is given by \cite{cove06}
\begin{eqnarray}
H_{z}
 &= -\frac{1}{\tau}\sum_{i,j,\alpha,\beta}P^\alpha_i W_{ij}^{\alpha\beta}\ln W_{ij}^{\alpha\beta}\nonumber\\
 &= -\sum_{i,\alpha,\beta\atop\alpha\neq \beta}P^\alpha_i w_{i}^{\alpha\beta}(\ln \tau+\ln w_{i}^{\alpha\beta}-1) \nonumber\\
 &\quad-\sum_{i,j\alpha\atop i\neq j}P^\alpha_iw_{ij}^{\alpha}(\ln \tau+\ln w_{ij}^\alpha-1)+\textrm{O}(\tau),
\label{entZ2}
\end{eqnarray}
The equality in the second line is convenient for the subsequent discussion where we will take the limit $\tau\to 0$. In general, 
a similar formula for the rates $H_{x}$ and $H_{y}$ in terms of the stationary distribution is not known and these Shannon entropy rates have to be calculated numerically \cite{bara13b,bara13a,jacq08,holl06,rold12}. 

A closely related quantity is the conditional Shannon entropy, which is defined as 
\begin{equation}
H(a_n|a_0^{n-1})\equiv \frac{1}{\tau}\sum_{a_0^n} P[a_0^n]\ln P[a_n|a_0^{n-1}].
\label{conddef}
\end{equation}
In the limit of $n\to \infty$ we have $H(a_n|a_0^{n-1})\to H_a$. Moreover, the conditional Shannon entropy decreases for increasing $n$: knowledge of a longer past 
decreases randomness \cite{cove06}. Therefore, the conditional Shannon entropies $H(x_n|x_0^{n-1})$ and $H(y_n|y_0^{n-1})$, which can be calculated in terms of the stationary probability distribution, provide an upper bound on the 
Shannon entropy rates $H_x$ and $H_y$, respectively. More precisely, it can be shown that for any finite $n$ and up to order $\tau$, the conditional Shannon entropies are given by \cite{bara13a}
\begin{equation}
H(x_n|x_0^{n-1})= -\sum_{i,\alpha}P^\alpha_i \sum_{\beta\neq\alpha} w_{i}^{\alpha\beta}(\ln \tau+ \ln \overline{w}^{\alpha\beta}-1)+\textrm{O}(\tau)
\label{condux}
\end{equation}   
and 
\begin{equation}
H(y_n|y_0^{n-1}) = -\sum_{i,\alpha}P^\alpha_i \sum_{j\neq i} w_{ij}^{\alpha}(\ln \tau+ \ln \overline{w}_{ij}-1)+\textrm{O}(\tau),
\label{conduy}
\end{equation}
where 
\begin{equation}
\overline{w}^{\alpha\beta}\equiv \sum_{i=1}^{\Omega_y}P(i|\alpha) w_{i}^{\alpha\beta}= \frac{1}{P^\alpha}\sum_{i=1}^{\Omega_y}P_i^\alpha w_{i}^{\alpha\beta}
\label{avgalpha}
\end{equation}
and
\begin{equation}
\overline{w}_{ij}\equiv \sum_{\alpha=1}^{\Omega_x}P(\alpha|i) w_{ij}^{\alpha}= \frac{1}{P_i}\sum_{\alpha=1}^{\Omega_x}P_i^\alpha w_{ij}^{\alpha},
\label{avgi}
\end{equation}
with $P_i= \sum_{\beta}P_i^\beta$ and $P^\alpha= \sum_{j}P_j^\alpha$.

The transfer entropy from $x$ to $y$ is defined as \cite{schr00}
\begin{equation}
T^n_{x\to y }\equiv H(y_n|y_0^{n-1})-H(y_n|x_0^{n-1},y_0^{n-1})\ge 0,
\label{transferxydef}
\end{equation}
where $H(y_n|x_0^{n-1},y_0^{n-1})\equiv -\frac{1}{\tau}\sum_{x_0^{n-1},y_0^n} P[x_0^{n-1},y_0^{n}]\ln P[y_n|x_0^{n-1},y_0^{n-1}]$.
It is the reduction on the conditional Shannon entropy of the $y$ process generated by knowing the $x$ process. In other words, it measures the 
dependence of $y$ on $x$ or the flow of information from $x$ to $y$. In the same way, the transfer entropy
from $y$ to $x$ is written as 
\begin{equation}
T^n_{y\to x }\equiv H(x_{n}|x_0^{n-1})-H(x_n|x_0^{n-1},y_0^{n-1})\ge 0.
\label{transferyxdef}
\end{equation}
The transfer entropy is in general not symmetric, i.e., $T^n_{x\to y }\neq T^n_{y\to x}$.

The conditional Shannon entropy $H(x_n|x_0^{n-1},y_0^{n-1})$ can be written as 
\begin{eqnarray}
H(x_n|x_0^{n-1},y_0^{n-1}) & = H(x_n|x_{n-1},y_{n-1})\nonumber\\
& = -\sum_{i,\alpha}P^\alpha_i\sum_{\beta\neq\alpha} w_{i}^{\alpha\beta}(\ln \tau+ \ln w_{i}^{\alpha\beta}-1)+\textrm{O}(\tau)
\label{condx}
\end{eqnarray}   
where the first  equality comes from the fact that the full process is Markovian and in the second equality we have performed the substitutions $x_{n-1}\to \alpha$, $y_{n-1}\to i$ and $x_n\to \beta$. Analogously, we obtain  
\begin{eqnarray}
H(y_n|x_0^{n-1},y_0^{n-1}) &= H(y_n|x_{n-1},y_{n-1})\nonumber\\
& = -\sum_{i,\alpha}P^\alpha_i\sum_{j\neq i} w_{ij}^{\alpha}(\ln \tau+ \ln w_{ij}^{\alpha}-1)+\textrm{O}(\tau).
\label{condy}
\end{eqnarray}

In this paper we are interested in the transfer entropy in the continuous time limit $\tau\to 0$, which is defined as  
\begin{eqnarray}
\mathcal{T}_{x\to y}\equiv
\lim_{\tau\to 0}\lim_{n\to \infty} T^n_{x\to y}=\lim_{\tau\to 0}\left(H_y+\sum_{i,\alpha}P^\alpha_i\sum_{j\neq i} w_{ij}^{\alpha}(\ln \tau+ \ln w_{ij}^{\alpha}-1)\right),\nonumber\\
\label{contransferxy}
\end{eqnarray}
where we used relation (\ref{condy}) in the second equality. The conditional Shannon entropies diverge as $\ln \tau$ for $\tau\to 0$ but the transfer entropy is well behaved in this limit. More clearly,
from formula (\ref{conduy}), the Shannon entropy rate $H_y= \lim_{n\to\infty} H(y_n|y_0^{n-1})$ diverges as $-\sum_{i,\alpha}P^\alpha_i\sum_{i\neq j} w_{ij}^{\alpha}\ln \tau$ in the limit $\tau\to 0$, canceling the 
$\ln \tau$ term in (\ref{contransferxy}). 

Moreover, as the conditional Shannon entropy (\ref{conduy}) for finite $n$ bounds the Shannon entropy rate $H_y$ from above, from equations (\ref{conduy}) and (\ref{condy}) 
we obtain an analytical upper bound on $\mathcal{T}_{x\to y}$, given by
\begin{equation}
\overline{\mathcal{T}}_{x\to y}= \sum_{i,\alpha}P_i^\alpha\sum_{j\neq i}w^\alpha_{ij}\ln \frac{w^\alpha_{ij}}{\overline{w}_{ij}}.
\label{upperxy}
\end{equation}
This result is similar to the analytical upper bound on the rate of mutual information (see \ref{a2}) we obtained in \cite{bara13b,bara13a}. Similarly, for the transfer entropy from $y$ to $x$, which is 
defined as $T^n_{y\to x }\equiv H(x_n|x_0^{n-1})-H(x_n|x_0^{n-1},y_0^{n-1})$, we obtain 
\begin{equation}
\overline{\mathcal{T}}_{y\to x}= \sum_{i,\alpha}P_i^\alpha\sum_{\beta\neq \alpha}w^{\alpha\beta}_{i}\ln\frac{w^{\alpha\beta}_{i}}{\overline{w}^{\alpha\beta}}.
\end{equation}
Furthermore, if there is a clear time scale separation, i.e., if the $x$ process is much faster than $y$, $w^{\alpha\beta}_i\gg w^{\alpha}_{ij}$, then $\mathcal{T}_{x\to y}\to\overline{\mathcal{T}}_{x\to y}$ \cite{bara13a}. Similarly, 
if the $y$ process is much faster  $\mathcal{T}_{y\to x}\to\overline{\mathcal{T}}_{y\to x}$.

While considering the discrete time case and taking the limit $\tau\to 0$ is convenient to calculate the upper bound (\ref{upperxy}), it is more efficient to consider continuous time trajectories with their waiting times to obtain 
the transfer entropy numerically. This issue and the relation between the transfer entropy and the rate of mutual information are discussed in \ref{a2}.

\section{Inequalities for transfer entropy}
\label{sec4}

\subsection{Integral fluctuation relation}

We consider a generic functional of the random variables $x^n_{n-1}$ and $y_0^{n}$, which is written as $\hat{F}[x_n,y_n;x_{n-1},y_0^{n-1}]$. The average of the functional is denoted by angular brackets, i.e.,
\begin{eqnarray}
\langle \hat{F}\rangle\equiv \sum_{x^n_{n-1},y_0^{n}}\hat{F}[x_n,y_n;x_{n-1},y_0^{n-1}] W[x_n,y_n|x_{n-1},y_{n-1}]P[x_{n-1},y_0^{n-1}],
\end{eqnarray}
where we used the Markov property $P[x_{n-1}^n,y_0^{n}]= W[x_n,y_n|x_{n-1},y_{n-1}]P[x_{n-1},y_0^{n-1}]$. Note that $W$ denotes a transition probability where the time
index $n$ is irrelevant. Particularly, we define the functionals
\begin{equation}
 \hat{\sigma}_x[x_n,y_n;x_{n-1},y_0^{n-1}]\equiv \ln \frac{W[x_n,y_n|x_{n-1},y_{n-1}]}{W[x_{n-1},y_{n}|x_{n},y_{n-1}]},
\end{equation}
and
\begin{equation}
\hat{T}_{x\to y}[x_n,y_n;x_{n-1},y_0^{n-1}]\equiv \ln \frac{P[x_{n-1}|y_0^{n-1}]}{P[x_{n}|y_0^{n-1}]}= \ln \frac{P[x_{n-1},y_0^{n-1}]}{P[x_{n},y_0^{n-1}]}.
\label{defMxyfr}
\end{equation}

We can prove the following integral fluctuation relation:
\begin{eqnarray}
& \langle \exp(-\hat{\sigma}_x-\hat{T}_{x\to y}) \rangle=\nonumber\\
& =  \sum_{x^n_{n-1},y_0^{n}}\left(\frac{W[x_{n-1},y_n|x_{n},y_{n-1}]}{W[x_{n},y_{n}|x_{n-1},y_{n-1}]}\frac{P[x_{n},y_0^{n-1}]}{P[x_{n-1},y_0^{n-1}]}\right)W[x_n,y_n|x_{n-1},y_{n-1}]P[x_{n-1},y_0^{n-1}]\nonumber\\
& =  \sum_{x^n_{n-1},y_0^{n}} W[x_{n-1},y_n|x_{n},y_{n-1}] P[x_{n},y_0^{n-1}]\nonumber\\
& =  \sum_{x_{n-1},y_n,x_n,y_0^{n-1}} W[x_{n-1},y_n|x_{n},y_{n-1}] P[x_{n},y_0^{n-1}]= \sum_{x_n,y_0^{n-1}} P[x_{n},y_0^{n-1}]= 1.
\label{fttransf}
\end{eqnarray}
This relation does not depend on the transition probabilities having the form (\ref{defrates}), therefore, it is valid also for systems that are not bipartite.  
Using Jensen's inequality we then obtain 
\begin{equation}
\langle \hat{\sigma}_x\rangle+\langle \hat{T}_{x\to y}\rangle\ge 0.
\label{jensentr}
\end{equation}
Finally, it is straightforward to show that 
\begin{equation}
\sigma_x= \frac{1}{\tau}\langle \hat{\sigma}_x\rangle.
\label{sigmaS}
\end{equation}
Furthermore, as we show in \ref{a3}, 
\begin{equation}
\mathcal{T}_{x\to y}= \lim_{\tau\to 0}\lim_{n\to \infty}\frac{1}{\tau}\langle \hat{T}_{x\to y}\rangle.
\label{transferT}
\end{equation}
Therefore,  inequality (\ref{jensentr}) implies 
\begin{equation}
\sigma_x+\mathcal{T}_{x\to y}\ge 0.
\label{inetransf}
\end{equation} 

A closely related fluctuation relation for causal networks has been recently obtained by Ito and Sagawa \cite{ito13}. To
obtain a fluctuation relation similar to (\ref{fttransf}) using the framework from \cite{ito13}, the stochastic trajectory of a bipartite system should be viewed
as a causal network with two connected rows, corresponding to the $x$ and $y$ processes.

Comparing our autonomous system, where there are no explicit measurements and feedback, with standard feedback driven systems,
the inequality (\ref{inetransf}) is analogous to the second law inequality for feedback driven systems as derived
in \cite{cao09}. As pointed out in \cite{saga12}, the quantity that bounds the extracted work in feedback driven systems is precisely 
the transfer entropy from the system to the controller performing the measurements, which is equivalent to $\mathcal{T}_{x\to y}$.     

\subsection{Comparison between $h_x$ and $\mathcal{T}_{x\to y}$}

We would like to compare the bounds $-\sigma_x\le h_x$ and $-\sigma_x\le \mathcal{T}_{x\to y}$ in order to assess which one is stronger.
By considering the average  $\langle\hat{\sigma}_x+ \hat{T}_{x\to y}\rangle$, we obtain the following inequality
\begin{eqnarray}
 &\left\langle
 \hat{\sigma}_x+\hat{T}_{x\to y}
 \right\rangle\nonumber\\
 &=
 \sum_{x^n_{n-1},y_{n-1}^{n}}\sum_{y_{0}^{n-2}}
 W[x_n,y_n|x_{n-1},y_{n-1}]P[x_{n-1},y_0^{n-1}]\nonumber\\
 &\quad\times\ln\left(\frac{W[x_{n},y_{n}|x_{n-1},y_{n-1}]}{W[x_{n-1},y_n|x_{n},y_{n-1}]}\frac{P[x_{n-1},y_0^{n-1}]}{P[x_{n},y_0^{n-1}]}\right)\nonumber\\
 &\ge
 \sum_{x^n_{n-1},y_{n-1}^{n}}
 W[x_n,y_n|x_{n-1},y_{n-1}]P[x_{n-1},y_{n-1}]\nonumber\\
 &\quad\times\ln\left(\frac{W[x_{n},y_{n}|x_{n-1},y_{n-1}]}{W[x_{n-1},y_n|x_{n},y_{n-1}]}\frac{P[x_{n-1},y_{n-1}]}{P[x_{n},y_{n-1}]}\right)
 =
 \langle \hat{\sigma}_x\rangle +\langle\hat {h}_{x}\rangle,
\label{ineq}
\end{eqnarray}
where we used the log sum inequality for the sum over $y_0^{n-2}$ and the definition 
\begin{equation}
\hat{h}_x[x_n,y_n;x_{n-1},y_0^{n-1}]\equiv \ln\frac{P[x_{n-1},y_{n-1}]}{P[x_{n},y_{n-1}]}.
\end{equation} 
By noting that $\langle\hat{h}_{x}\rangle/\tau= h_x+\textrm{O}(\tau)$, we obtain that the inequality (\ref{ineq}) in the limit $\tau\to0$ implies 
\begin{equation}
 h_x\le\mathcal{T}_{x\to y}.
\label{hxmxy}
\end{equation}
Therefore, $h_x$ provides a better bound on $-\sigma_x$ than the transfer entropy $\mathcal{T}_{x\to y}$. 

The above inequality can also be seen in a different way. In general, we do not know an analytical expression for the transfer entropy in terms of 
the stationary distribution. With the upper bound (\ref{upperxy}) and the inequality (\ref{hxmxy})
we find
\begin{equation}
 h_x\le\mathcal{T}_{x\to y}\le\overline{\mathcal{T}}_{x\to y}.
\label{upperlower}
\end{equation}

Summarizing inequalities (\ref{eq:sx_hy_sy}) and (\ref{upperlower}), we obtain
\begin{equation}
 -\sigma_x\le h_x\le\cases{\sigma_y\\\mathcal{T}_{ x\to y}\le\overline{\mathcal{T}}_{ x\to y}.}
\label{summine}
\end{equation}
Knowing now that $h_x$ is the best bound on $-\sigma_x$ we also would like to investigate whether there is a unique relation between $\sigma_y$ and the transfer entropy $\mathcal{T}_{x\to y}$. The following
example will demonstrate that there is no such inequality.


\section{Four state system}
\label{sec5}

\begin{figure}%
 \centering%
\includegraphics{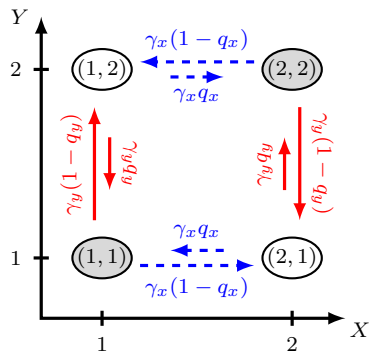}
 \caption{Four state system.}
\label{fig:4state_system}%
\end{figure}%

To illustrate our results we consider the simplest bipartite system which is a four states model. The transition rates are defined  in Fig. \ref{fig:4state_system}.
The parameters $\gamma_x$ and $\gamma_y$ set the timescales of the $x$ and $y$ transitions, respectively. We consider the case where $q_y\le q_x\le1/2$ so that the
probability current runs in the clockwise direction. In this case, $\sigma_y$ and $-\sigma_x$ are both positive. 

We can interpret the model of Fig. \ref{fig:4state_system} as follows. We consider two coupled proteins $x$ and $y$ that each can be in an 
inactive or active state, represented by $1$ and $2$, respectively. A chemical reaction, with chemical potential difference $\Delta \mu_x\ge0$, drives the
transitions of the $x$ protein favoring the states $(1,2)$ and $(2,1)$, where the proteins are in different configurations. Local detailed balance is then written as
\begin{equation}
\ln\frac{1-q_x}{q_x}= \Delta \mu_x.
\end{equation}
Another chemical reaction drives the $y$ transitions, also favoring anti-alignment of the proteins, implying in the local detailed balance relation
\begin{equation}
\ln\frac{1-q_y}{q_y}= \Delta \mu_y. 
\end{equation}   
The condition $q_y\le q_x\le1/2$ reads $\Delta \mu_y\ge \Delta \mu_x\ge0$. In this case the chemical reaction driving the $y$ transitions feeds work into the
system at a rate $\sigma_y$ and the system does work against the chemical reaction driving the $x$ transitions at a rate $-\sigma_x$.  

Explicitly, the stationary current is given by
\begin{equation}
J= P_1^2\gamma_xq_x-P_1^1\gamma_x(1-q_x),
\end{equation}  
with the stationary probabilities $2P_1^1=2P^2_2=(\gamma_xq_x+\gamma_yq_y)/(\gamma_x+\gamma_y)$ and $2P_1^2=2P^1_2=1-2P_1^1$. Moreover, the rate of extracted work is given by 
\begin{equation}
 -\sigma_x=2J\ln\left(\frac{1-q_x}{q_x}\right)\ge0,
\end{equation}
and the rate of energy input is
\begin{equation}
 \sigma_y=2J\ln\left(\frac{1-q_y}{q_y}\right)\ge0.
\end{equation}
The rate of entropy reduction of the subsystem $x$ due to its coupling to $y$ is  
\begin{equation}
 h_x=2J\ln\left(\frac{\gamma_x(1-q_x)+\gamma_y(1-q_y)}{\gamma_xq_x+\gamma_yq_y}\right)\ge0.
\end{equation}
The upper bound of the transfer entropy reads 
\begin{equation}
\overline{\mathcal{T}}_{x\to y}=2P_1^2\gamma_yq_y\ln\frac{q_y}{r_y}+2P_1^1\gamma_y(1-q_y)\ln\frac{1-q_y}{r_y},
\end{equation}
where $r_y\equiv 2P_1^2q_y+2P_1^1(1-q_y)$. An analytical expression for the transfer entropy $\mathcal{T}_{x\to y}$ is not known but we can determine it numerically.

\begin{figure}
\centering
\includegraphics{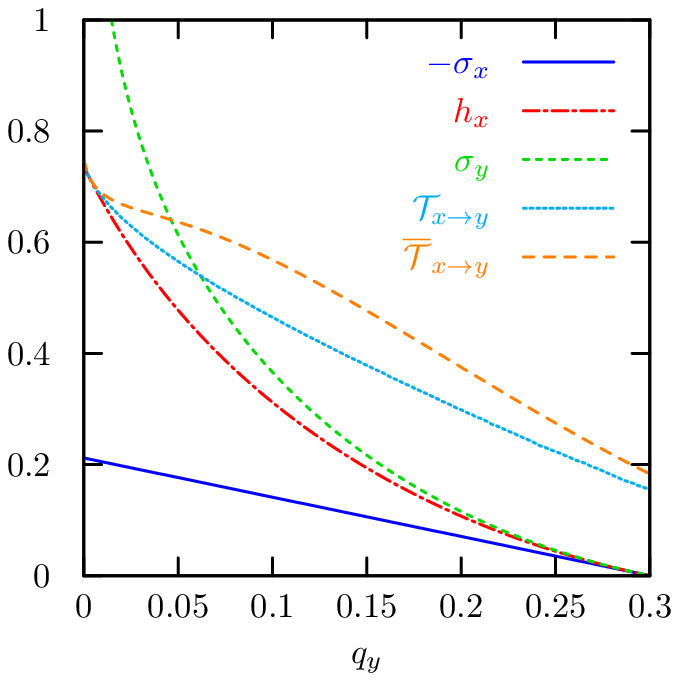}
\includegraphics{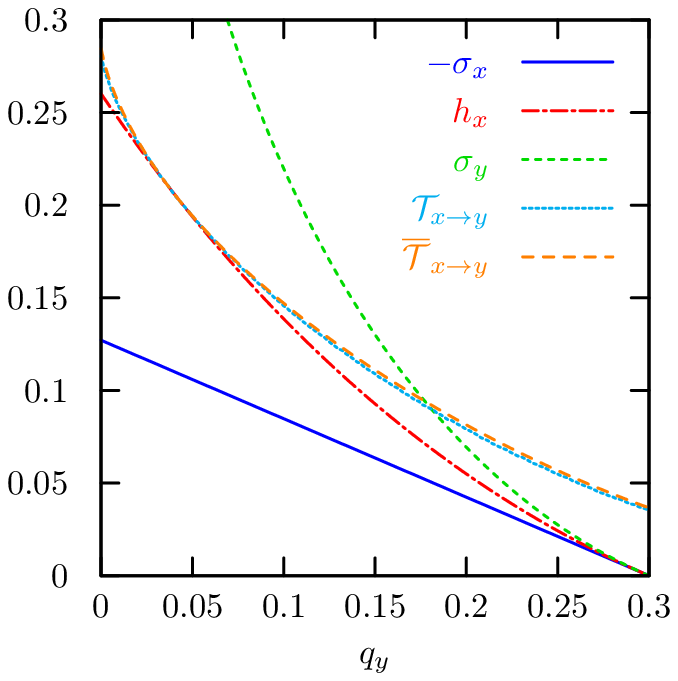}
 \caption{The rates of extracted work $-\sigma_x$ and the bounds $h_x$, $\sigma_y$, $\mathcal{T}_{x\to y}$, $\overline{\mathcal{T}}_{x\to y}$ as a function of $q_y$
for $q_x=0.3$, $\gamma_x=1$, and $\gamma_y=5$ (left panel), $\gamma_y=1$ (right panel) for the network shown in Fig. \ref{fig:4state_system}. The transfer entropy $\mathcal{T}_{x\to y}$ is calculated using the numerical method from \cite{bara13a}, as
explained in \ref{a2}.}
 \label{fig:ey0.3_ax5_ay1}
\end{figure}

In Fig. \ref{fig:ey0.3_ax5_ay1} we compare the three different bounds on the extracted work  $-\sigma_x$. Besides the illustration of inequalities (\ref{summine}), we also can see that  $\mathcal{T}_{x\to y}$
approaches $\overline{\mathcal{T}}_{x\to y}$ as the $y$ process becomes slower, as discussed in Sec. \ref{sec3}. The main result we obtain from these plots is
the crossing of the transfer entropy $\mathcal{T}_{x\to y}$ and $\sigma_y$, with the input $\sigma_y$ being smaller near equilibrium $q_x=q_y$ and  larger in the far from equilibrium limit $q_y\to 0$.


\section{Conclusion}
\label{sec6}

We have studied a series of second law like inequalities valid for bipartite systems. Besides the standard entropy production (\ref{sdef1}), the
entropy production of a subsystem (\ref{totx}) and the inequalities involving transfer entropy (\ref{inetransf}) and (\ref{hxmxy}) have been analyzed.
Moreover, inspired by the fluctuation relation recently obtained by Ito and Sagawa \cite{ito13} we have obtained the fluctuation relation (\ref{fttransf}),
which in the continuous time limit leads to the inequality involving transfer entropy. From the summary of the inequalities (\ref{summine}) we have obtained 
that $h_x$, the rate of entropy reduction of $x$ due to the coupling with $y$, provides the best bound on $-\sigma_x$. As a particularly interesting interpretation,
we have shown that a bipartite system provides a transparent realization of Maxwell's demon, with an integrated description of the subsystem and 
demon being easily accessible through the standard entropy production. 

Analyzing a simple four state model we have shown that the transfer entropy $\mathcal{T}_{x\to y}$ can be larger than the entropy rate proportional to the heat dissipated by the 
demon $\sigma_y$. While the crossing between $\mathcal{T}_{x\to y}$ and $\sigma_y$ has been obtained for a specific model we conjecture it to be more general because it depends on two general properties:
$\mathcal{T}_{x\to y}$ being not zero in equilibrium, where $\sigma_y=0$, and $\mathcal{T}_{x\to y}$ being finite when a $y$ transition rate goes to zero, where $\sigma_y$ diverges.
Furthermore, as transfer entropy is generally not zero in equilibrium it should be useless as a bound on $-\sigma_x$ near equilibrium, e.g., in the linear response regime.   

It is interesting to compare the present work with \cite{bara14}, where a simplified version of the Mandal and Jarzynski model \cite{mand13} for a tape interacting with a thermodynamic system was analyzed.
In \cite{bara14} the entropy (or information) reservoir is a tape composed of a sequence of bits while here it is the $y$ subsystem.  Moreover, a series of inequalities similar to (\ref{summine}) have been obtained in \cite{bara14},
with the Shannon entropy difference of the tape providing the best bound on the extracted work, as is the case of $h_x$ here. Likewise, the mutual information between the tape and the system crosses the full input of work to
reset the tape (see also \cite{horo13}), corresponding to the crossing of $\mathcal{T}_{x\to y}$ and $\sigma_y$ here.

Summarizing, the series of inequalities studied here, and the methods to calculate quantities like the rate of mutual information and transfer entropy that we have developed in \cite{bara13a} form a solid
theoretical framework for bipartite systems, which constitute an important class of Markov processes. Among possible applications of our results, investigating bipartite models for cellular sensing
is an interesting direction for future work.

{\noindent \textbf{Acknowledgements}}\newline Support by the ESF through the network EPSD  is gratefully acknowledged.

\appendix

\section{Fluctuation relation for the total entropy of the subsystem}
\label{a1}

The full Markovian stochastic trajectory from time $0$ to $T$ is denoted by 
$z(t)_0^T=(z_0,\tau_0;z_1,\tau_1;\ldots;z_N,\tau_N)$, where the waiting times fulfill $\tau_0+\tau_1+\ldots+\tau_N=T$ and $N$ is the number of jumps in the trajectory.
The probability density of a trajectory is written as
\begin{equation}
P[z(t)_0^T]= P(z_0)\prod_{n=0}^{N-1}w_{z_nz_{n+1}}\prod_{n=0}^{N}\exp(-\lambda_{z_n}\tau_n)
\label{pathprob}
\end{equation}
where $P(z_0)$ is the initial distribution, $w_{z_nz_{n+1}}$ denotes the transition rate from $z_n$ to $z_{n+1}$ defined in (\ref{defrates2}) and $\lambda_{z_n}\equiv\sum_zw_{z_nz}$ is the escape rate.

In order to obtain the fluctuation relation leading to (\ref{eq:hx_def}) we consider the modified transition rates  
\begin{equation}
u_{ij}^{\alpha\beta}\equiv\left\{
\begin{array}{ll} 
w^{\beta\alpha}_iP_i^\beta/P_i^\alpha & \quad \textrm{if $i=j$ and $\alpha\neq\beta$}, \\
w^{\alpha}_{ij} & \quad  \textrm{if $i\neq j$ and $\alpha=\beta$},\\
0 & \quad \textrm{if $i\neq j$ and $\alpha\neq\beta$}. 
\end{array}\right.\,
\end{equation}
and denote the path probability (\ref{pathprob}) obtained with these modified transition rates by $P^{\dagger}[z(t)_0^T]$, where the initial probability is also $P(z_0)$. The escape rates for $u$
are written as $\psi_{z_n}\equiv\sum_zu_{z_nz}$. Furthermore, 
we define the functionals 
\begin{equation}
\Delta H_x[z(t)_0^T]\equiv \sum_{n=0}^{N-1} \delta_{y_n,y_{n+1}}\ln\frac{P_{y_{n}}^{x_{n}}}{P_{y_{n+1}}^{x_{n+1}}}=\sum_{n=0}^{N-1} \ln\frac{P_{y_{n}}^{x_{n}}}{P_{y_{n}}^{x_{n+1}}},
\label{hxfunc}
\end{equation}
\begin{equation}
\Delta S_x[z(t)_0^T]\equiv \sum_{n=0}^{N-1} \delta_{y_n,y_{n+1}}\ln\frac{w_{z_nz_{n+1}}}{w_{z_{n+1}z_{n}}},
\end{equation}
where $\delta_{y_n,y_{n+1}}$ is the Kronecker delta function. In the limit $T\to\infty$ we have $\langle\Delta H_x\rangle/T\to h_x$ and $\langle\Delta S_x\rangle/T\to \sigma_x$, where the angular brackets
here denote an integral over all stochastic paths (note that this is different from Sec. \ref{sec4}). 

The usual ratio of path probabilities is then given by
\begin{equation}
\frac{P^{\dagger}[z(t)_0^T]}{P[z(t)_0^T]}= \exp(-\Delta H_x[z(t)_0^T]-\Delta S_x[z(t)_0^T]+\Lambda_x[z(t)_0^T]),
\label{ratioprob}
\end{equation}
where the functional $\Lambda_x[z(t)_0^T]\equiv \sum_{n=0}^{N}(\lambda_{z_n}-\psi_{z_n})\tau_n$ comes from the fact that the escape rates of $w$ and $u$ are different. 
Using standard methods \cite{seif12}, relation (\ref{ratioprob}) implies  
\begin{equation}
\langle \exp\left(-\Delta H_x[z(t)_0^T]-\Delta S_x[z(t)_0^T]+\Lambda_x[z(t)_0^T]\right) \rangle=1.
\end{equation}
From Jensen's inequality and
\begin{equation}
\langle\Lambda_x\rangle/T\to \sum_{i,\alpha}P_i^{\alpha}\sum_{\beta\neq\alpha}(w^{\alpha\beta}_i-w^{\beta\alpha}_iP_i^{\beta}/P_i^{\alpha})=0, 
\end{equation}
we obtain the second law for the subsystem $x$ (\ref{totx}).

Let us make the following remarks. One could consider an $x$ transition from $\alpha$ to $\beta$ as dependent on
time due to changes in the variable $i$ in the transition rates $w_{i}^{\alpha\beta}$. Within this view, the rates $u$ corresponds to a sort of ``adjoint'' dynamics
and relation (\ref{ratioprob}) is similar to the ratio of probabilities involving the forward adjoint trajectory in the fluctuation relation for the house keeping entropy derived in \cite{spec05a}.
Furthermore, denoting by $N_y$ the number of jumps for which the variable $y$ changes  and considering the interval between two $y$ jumps $[n_y,n_y+1]$ we write 
\begin{equation}
\delta H_x(n_y)= \ln\frac{P^{x_i}_j}{P^{x_f}_j},
\end{equation} 
where $x_i$ is the $x$ state at the time of the jump $n_y$, $x_f$ is the $x$ state at the time of the jump $n_y+1$ and $j$ is the $y$ state in the time interval between the jumps $n_y$ and $n_y+1$. The functional (\ref{hxfunc}) can then 
be written as
\begin{equation}
\Delta H_x[z(t)_0^T]\equiv \sum_{n_y=0}^{N_y-1} \delta H_x(n_y).
\end{equation} 
In this form it becomes clear that the rate $h_x= \langle\Delta H_x\rangle/T$ in the large $T$ limit is the rate of the entropy reduction of the subsystem $x$ due to the subsystem $y$ dynamics.

\section{Transfer entropy in continuous time} 
\label{a2}

Using the notation of \ref{a1}, the continuous time Shannon entropy rate is given by \cite{dumi88}
\begin{eqnarray}
\mathcal{H}_{z}&\equiv -\lim_{T\to \infty}\frac{1}{T}\sum_{z(t)_0^T}P[z(t)_0^T]\ln P[z(t)_0^T]\nonumber\\
&=\sum_{i,\alpha}P^\alpha_i\sum_{\beta\neq\alpha}w_{i}^{\alpha\beta}(\ln w_{i}^{\alpha\beta}-1) +\sum_{i,\alpha}P^\alpha_i\sum_{j\neq i}w_{ij}^{\alpha}(\ln w_{ij}^\alpha-1).
\label{entratez	}
\end{eqnarray}
If we compare this formula with (\ref{entZ2}) we see that the continuous time entropy rate does not show the $\ln \tau$ divergence. 

The coarse grained trajecories are written as  $x(t)_0^T=(x_0,\tau_0^x;x_1,\tau_1^x;\ldots;x_{N_x},\tau_{N_x}^x)$ and $y(t)_0^T=(y_0,\tau_0^y;y_1,\tau_1^y;\ldots;y_{N_y},\tau_{N_y}^y)$, where
$\tau_0^x+\tau_1^x+\ldots+\tau_{N_x}^x=\tau_0^y+\tau_1^y+\ldots+\tau_{N_y}^y=T$ and $N_x$ ($N_y$) is the number of jumps where the $x$ ($y$) variable changes. Due to the bipartite nature of the transition rates $N=N_x+N_y$. 
The continuous time Shannon entropy rate of the $y$ process is defined as
\begin{equation}
\mathcal{H}_{y}\equiv -\lim_{T\to \infty}\frac{1}{T}\sum_{y(t)_0^T}P[y(t)_0^T]\ln P[y(t)_0^T].
\end{equation}
In contrast to $\lim_{\tau\to 0}H_y$, this continuous time Shannon entropy rate does not have the divergent term proportional to $\ln\tau$. Hence, the transfer entropy (\ref{contransferxy}) can be written as 
\begin{equation}
\mathcal{T}_{x\to y}= \mathcal{H}_y-\sum_{i,\alpha}P^\alpha_i\sum_{j\neq i} w_{ij}^{\alpha}(\ln w_{ij}^{\alpha}-1).
\end{equation}
Therefore, to calculate the transfer entropy we just have to obtain the non-Markovian entropy rate $\mathcal{H}_y$: this can be achieved by using a numerical method to estimate Shannon
entropy for non-Markovian continuous time processes developed in \cite{bara13a}.  

We can apply the same procedure to the transfer entropy from $y$ to $x$, which is written as
\begin{equation}
\mathcal{T}_{y\to x}= \mathcal{H}_x-\sum_{i,\alpha}P^\alpha_i\sum_{\beta\neq\alpha} w_{i}^{\alpha\beta}(\ln w_{i}^{\alpha\beta}-1).
\label{eq:Myx_cont}
\end{equation}
Finally we would like to point out the relation
\begin{equation}
\mathcal{I}\equiv \mathcal{H}_x+\mathcal{H}_y-\mathcal{H}_z=\mathcal{T}_{x\to y}+\mathcal{T}_{y\to x},
\end{equation}
between the rate of mutual information $\mathcal{I}$ and the transfer entropy. The rate of mutual information measures how correlated the $x$ and $y$ processes are, without any specific direction. Obviously,
the lower bound on transfer entropy (\ref{hxmxy}) can be extended to a lower bound on the rate of mutual information, i.e., $\mathcal{I}\ge |h_x|$.

\section{Proof of relation (\ref{transferT})} 
\label{a3}

We start by rewriting  (\ref{defMxyfr}) in the from
\begin{equation}
\frac{1}{\tau}\langle \hat{T}_{x\to y}\rangle = H(x_n|y_0^{n-1})-H(x_{n-1}|y_0^{n-1}).
\end{equation}
We are interested in the limit $n\to \infty$ and for $n$ large enough we can substitute $H(x_{n-1}|y_0^{n-1})$ by $H(x_{n}|y_0^{n})$, leading to
\begin{equation}
\frac{1}{\tau}\langle \hat{T}_{x\to y}\rangle = H(x_n|y_0^{n-1})-H(x_{n}|y_0^{n})= H(y_n|y_0^{n-1})-H(y_{n}|x_n,y_0^{n-1}).
\end{equation}
Hence, from the definition (\ref{contransferxy}), in order to prove (\ref{transferT}) we have to show that
\begin{equation}
H(y_{n}|x_n,y_0^{n-1})= -\sum_{i,j,\alpha\atop i\neq j}P^\alpha_i w_{ij}^{\alpha}(\ln \tau+ \ln w_{ij}^{\alpha}-1)+\textrm{O}(\tau).
\label{eq:A_Hvar1}
\end{equation}
Comparing this with (\ref{condx}), which reads
\begin{equation}
H(y_{n}|x_{n-1},y_0^{n-1})= -\sum_{{i,j,\alpha\atop i\neq j}}P^\alpha_i w_{ij}^{\alpha}(\ln \tau+ \ln w_{ij}^{\alpha}-1)+\textrm{O}(\tau),
\label{eq:A_Hvar2}
\end{equation}
it is then analogous to demonstrate that
\begin{equation}
\left\langle\ln\left\{\frac{P[y_n|x_n,y_0^{n-1}]}{P[y_n|x_{n-1},y_0^{n-1}]}\right\}\right\rangle=\textrm{O}(\tau^2).
\label{eq:Otau2}
\end{equation}
The term inside the logarithm can be written as
\begin{eqnarray}
& \frac{P[y_n|x_n,y_0^{n-1}]}{P[y_n|x_{n-1},y_0^{n-1}]}= \frac{P[y_n,x_n,y_0^{n-1}]}{P[x_n,y_0^{n-1}]P[y_n|x_{n-1},y_{n-1}]}\label{eq:BracketDots}\\
&=
\frac{
 \sum_{\tilde x_{n-1}}
 W[x_{n},y_{n}|\tilde{x}_{n-1},y_{n-1}]
 P[\tilde{x}_{n-1},y_0^{n-1}]
}{
 \sum_{\tilde x_{n-1},\tilde y_{n}}
 W[x_{n},\tilde{y}_{n}|\tilde{x}_{n-1},y_{n-1}]P[\tilde{x}_{n-1},y_0^{n-1}]
 \sum_{\tilde{x}_n}W[\tilde{x}_n,y_n|x_{n-1},y_{n-1}]
}
.\nonumber
\end{eqnarray}
In the following we consider three cases.

First, we consider $y_n\neq y_{n-1}$, which implies $x_n=x_{n-1}$. Equation (\ref{eq:BracketDots}) takes the form
\begin{eqnarray}
\label{firsteq} 
&\frac{P[y_n|x_n,y_0^{n-1}]}{P[y_n|x_{n-1},y_0^{n-1}]}\\
&=
\frac{
 W[x_{n-1},y_{n}|x_{n-1},y_{n-1}]
 P[x_{n-1},y_0^{n-1}]
}{
 \sum_{\tilde x_{n-1},\tilde y_{n}}
 W[x_{n-1},\tilde{y}_{n}|\tilde{x}_{n-1},y_{n-1}]P[\tilde{x}_{n-1},y_0^{n-1}]
 W[x_{n-1},y_n|x_{n-1},y_{n-1}]
}\nonumber\\
&=
\frac{
1
}{
\sum_{\tilde x_{n-1},\tilde y_{n}}
 W[x_{n-1},\tilde{y}_{n}|\tilde{x}_{n-1},y_{n-1}]\frac{P[\tilde{x}_{n-1},y_0^{n-1}]}{P[x_{n-1},y_0^{n-1}]}
}=1+\textrm{O}(\tau),\nonumber
\end{eqnarray}
where in the last equality we used $W[x_{n-1},\tilde{y}_{n}|\tilde{x}_{n-1},y_{n-1}]=1+\textrm{O}(\tau)$ for $\tilde{x}_{n-1}= x_{n-1}$ and $\tilde{y}_{n}= y_{n-1}$, 
and  $W[x_{n-1},\tilde{y}_{n}|\tilde{x}_{n-1},y_{n-1}]=\textrm{O}(\tau)$ otherwise.

Second, we take $x_n \neq x_{n-1}$ implying $y_n=y_{n-1}$.
The term (\ref{eq:BracketDots}) is now written as
\begin{eqnarray}
 &\frac{P[y_n|x_n,y_0^{n-1}]}{P[y_n|x_{n-1},y_0^{n-1}]}\nonumber\\
&=
 \frac{
 \sum_{\tilde x_{n-1}}
 W[x_{n},y_{n-1}|\tilde{x}_{n-1},y_{n-1}]
 P[\tilde{x}_{n-1},y_0^{n-1}]
}{
 \sum_{\tilde x_{n-1},\tilde y_{n}}
 W[x_{n},\tilde{y}_{n}|\tilde{x}_{n-1},y_{n-1}]P[\tilde{x}_{n-1},y_0^{n-1}]
 \sum_{\tilde{x}_n}W[\tilde{x}_n,y_{n-1}|x_{n-1},y_{n-1}]}\nonumber\\
&=1+\textrm{O}(\tau).
\label{secondeq}
\end{eqnarray}
where we used $\sum_{\tilde{x}_n}W[\tilde{x}_n,y_{n-1}|x_{n-1},y_{n-1}]=1+\textrm{O}(\tau)$.

Third, we consider  $x_{n}=x_{n-1}$ and $y_{n}=y_{n-1}$.
It is convenient to define $L$ through the equality $W[x,y|x',y']=\delta_{x,x'}\delta_{y,y'}+\tau L[x,y|x',y']$, where $L$ is
the stochastic matrix corresponding to the transiton rates (\ref{defrates2}). The term (\ref{eq:BracketDots}) now becomes
\begin{eqnarray}
 \frac{P[y_n|x_n,y_0^{n-1}]}{P[y_n|x_{n-1},y_0^{n-1}]}
&=
\frac{
 1+\sum_{\tilde x_{n-1}}
 \tau L[x_{n-1},y_{n-1}|\tilde{x}_{n-1},y_{n-1}]
 \frac{P[\tilde{x}_{n-1},y_0^{n-1}]}{P[x_{n-1},y_0^{n-1}]}
}{
1+
 \sum_{\tilde{x}_{n-1},\tilde y_{n}}
 \tau L[x_{n-1},\tilde{y}_{n}|\tilde{x}_{n-1},y_{n-1}]\frac{P[\tilde{x}_{n-1},y_0^{n-1}]}{P[x_{n-1},y_0^{n-1}]}
}\nonumber\\
&\quad\times\left(1+\sum_{\tilde{x}_n}\tau L[\tilde{x}_n,y_{n-1}|x_{n-1},y_{n-1}]\right)^{-1}.
\label{eq:Average3rdCase}
\end{eqnarray}

Finally the quantity (\ref{eq:Otau2}) can be separated into three contributions: one for $y_n\neq y_{n-1}$, the second for $x_n\neq x_{n-1}$, and the third
for $x_{n}=x_{n-1}$ and $y_{n}=y_{n-1}$. Considering equations (\ref{firsteq}) and (\ref{secondeq}), we see that the first two contributions give $\textrm{O}(\tau^2)$, leading to
\begin{eqnarray}
  &\left\langle\ln\frac{P[y_n|x_n,y_0^{n-1}]}{P[y_n|x_{n-1},y_0^{n-1}]}\right\rangle\\
  &=\sum_{x_{_n-1},y_0^{n-1}}\left(1+\tau L[x_{_n-1},y_{n-1}|x_{_n-1},y_{n-1}]\right)P[x_{n-1},y_0^{n-1}]\ln \frac{P[y_n|x_n,y_0^{n-1}]}{P[y_n|x_{n-1},y_0^{n-1}]}+\textrm{O}(\tau^2).\nonumber
\end{eqnarray}
Using equation (\ref{eq:Average3rdCase}) we obtain
\begin{eqnarray}
 &\left\langle\ln\frac{P[y_n|x_n,y_0^{n-1}]}{P[y_n|x_{n-1},y_0^{n-1}]}\right\rangle\\
 &=
 \sum_{\tilde x_{n-1},x_{n-1},y_0^{n-1}}
 \tau L[x_{n-1},y_{n-1}|\tilde{x}_{n-1},y_{n-1}]
 P[\tilde{x}_{n-1},y_0^{n-1}]\nonumber\\
 &\quad-
 \sum_{\tilde{x}_{n-1},x_{n-1},\atop\tilde y_{n},y_0^{n-1}}
 \tau L[x_{n-1},\tilde{y}_{n}|\tilde{x}_{n-1},y_{n-1}]P[\tilde{x}_{n-1},y_0^{n-1}]
 \nonumber\\
 &\quad-\sum_{\tilde x_{n},x_{n-1},y_0^{n-1}}\tau L[\tilde{x}_n,y_{n-1}|x_{n-1},y_{n-1}]P[x_{n-1},y_0^{n-1}]
+\textrm{O}(\tau^2)=\textrm{O}(\tau^2),\nonumber
 \end{eqnarray}
where, from  $\sum_{x,y}L[x,y|x',y']=0$ for all $x',y'$, the term in the third line is zero and the terms in the second and fourth lines cancel.
This concludes the proof of (\ref{eq:Otau2}), which implies (\ref{transferT}).

\printbibliography

\end{document}